# Pressure-Induced Re-entrant transition in NbS$_3$ Phases: Combined Raman Scattering and X-ray Diffraction Study

M. Abdel-Hafiez[1,2,3,4*] R. Thiyagarajan[1], A. Majumdar[4], R. Ahuja[4], W. Luo[4], A. N. Vasiliev[3,5,6], A.A. Maarouf[7], S.G. Zybtsev,[8] V. Ya. Pokrovskii,[8] S. V. Zaitsev-Zotov,[8] V.V. Pavlovskiy,[8] Woei Wu Pai,[9,10] W. Yang[1,11], and L.V. Kulik[12]

[1]*Center for High Pressure Science and Technology Advanced Research, Shanghai 201203, China*
[2]*Lyman Laboratory of Physics, Harvard University, Cambridge, Massachusetts 02138, USA*
[3]*National University of Science and Technology "MISiS", Moscow 119049, Russia*
[4]*Department of Physics and Astronomy, Box 516, Uppsala University, Uppsala, SE-75120, Sweden*
[5]*National Research South Ural State University, Chelyabinsk 454080, Russia*
[6]*Moscow State University, Moscow 119991, Russia*
[7]*Department of Physics, IRMC, Imam Abdulrahman Bin Faisal University, Saudia Arabia*
[8]*Institute of Radioengineering and Electronics, RAS, Moscow 125009, Russia*
[9]*Center for Condensed Matter Sciences, National Taiwan University, Taipei 106, Taiwan*
[10]*Department of Physics, National Taiwan University, Taipei, Taiwan 10610*
[11]*Center of Matter at Extreme Conditions, Florida International University, Miami, Florida 33199, USA*
[12]*Institute of Solid State Physics, Russian Academy of Sciences, 142432 Russia*

We report the evolution of charge density wave states under pressure for two NbS$_3$ phases – triclinic (phase I) and monoclinic (phase II) at room temperature. Raman and X-ray diffraction (XRD) techniques are applied. The x-ray studies on the monoclinic phase under pressure show a compression of the lattice at different rates below and above ~7 GPa but without a change in space group symmetry. The Raman spectra of the two phases evolve similarly with pressure; all peaks almost disappear in the ~ 6 – 8 GPa range, indicating a transition from an insulating to a metallic state, and peaks at new positions appear above 8 GPa. The results suggest suppression of the ambient charge-density waves and their subsequent recovery with new orderings above 8 GPa.

## I. INTRODUCTION
### A. Effects of pressure on low-dimensional compounds

Fermi surface nesting in quasi-1D materials such as transitional metal trichalchogenides (MX$_3$, M: transitional metal, X: chalcogen atoms) is an important condition of charge-density waves (CDWs) formation [1,2]. Consequently, by manipulating the anisotropy of such materials, one can modify the CDWs configuration to suppress them. New correlated electronic states such as superconductivity, new CDWs or other-type electronic orderings can also emerge. Anisotropy can be altered by internal or external perturbations. For example, the CDW state of the RTe$_3$ (R: rare-earth elements) series is progressively suppressed by lattice chemical compression, i.e., substituting from R=La to R=Tm, and finally becomes unstable [3]. Alternatively, the materials can be subjected to external uniaxial or hydrostatic pressure to alter its anisotropy. For quasi-1D MX$_3$, external pressure reduces the distance between the quasi-1D chains, therefore impairing the Fermi surface nesting conditions and resulting in an increase in the effective dimensionality of the quasi-1D conductors. Hydrostatic pressure ($P$) results in suppression of CDW and in some cases induce superconductivity at the highest $P$ values [2]. Usually, the CDW transition temperatures, $T_P$, decrease with pressure [2,4], with the exceptions of the monoclinic phase of TaS$_3$, whose upper CDW transition temperature slightly grows up to

$P \sim 4$ GPa, and for (TaSe$_4$)$_2$I, whose $T_P$ is also likely to grow at pressures up to 1 GPa. Another exception is a 2D chain semiconducting compound ZrTe$_3$, where enhancement of the CDW transition temperature is reported for the moderate $P$, up to 2 GPa [5]. In this remarkable compound, the CDW distortion is normal to the highly conducting *b*-axis with the wave-vector $q = (1/14,0,1/3)$. Thus, for a CDW directed perpendicular to the quasi-1D chains, an effective reduction of dimensionality can even increase the CDW transition temperature.

The CDWs can show not only suppression but also fine transformations under pressure. The CDW configurations are more diverse in 2D and their rearrangements under pressure are better studied in layered compounds [6,7,8]. Nonetheless, in 1D cases, the CDWs can also change their configuration under pressure. Orthorhombic TaS$_3$ is an example. In refs. [2,4] one can notice a feature in $T_P(P)$ at $P \sim 2$ GPa, which could reflect a formation of a new CDW. This assumption is supported by observation of a 1$^{st}$ order-like transition of the CDW into a new state under a critical strain [9,10,11], taking in view that the effect of pressure is usually close to that of uniaxial strain (see [12], e.g.). An interplay of two CDWs under pressure, one of which is suppressed and the other – stimulated, is suggested for another quasi-1D compound, SmNiC$_2$ [13]. Apart from dimensionality, $P$ can also alter the electronic structure of the initial phase, e.g., transferring a certain number of charge carriers to charge-reservoir layers [14,15,16]. Given the diverse responses of CDW in low-dimensional materials under pressure, we undertake a pressure-dependent study on a material, NbS$_3$, which uniquely shows multiple polymorphs and also multiple CDW transitions [12].

### B. Two phases of NbS$_3$

NbS$_3$ has been synthesized in a number of structure types [2,17], among which the two phases, triclinic NbS$_3$ (NbS$_3$-I) and monoclinic NbS$_3$ (NbS$_3$-II), have been studied in detail. Being a typical representative of the MX$_3$ family of the quasi-1D conductors, NbS$_3$-II is remarkable for its three CDWs, each of which can slide [12,18]. The semiconducting triclinic phase, NbS$_3$-I, shows dimerization along the chains (*b*-axis). The dimerization, which is stable upon temperature variations [2,19], results in the lowered symmetry of the crystal structure. It is not surprising that transport properties of the two phases are strongly different. NbS$_3$-II shows an increase of resistance ($R$) with a lower temperature ($T$) and is most pronounced across the CDW transition temperatures at $T_{P0}$(CDW-0) = 450-475 K, $T_{P1}$(CDW-1) = 360 K and $T_{P2}$(CDW-2) = 150 K. The $T_{P2}$ transition is typical of the low-Ohmic samples, which appear to be a "subphase" of phase II [12]. The batch studied in this paper contains both low- and high-Ohmic samples. More details of the properties of NbS$_3$-II can be found in [12].

Despite the fact that in NbS$_3$-II the electrons at room temperature are gapped already by two coexisting CDWs (CDW-0 and CDW-1), the resistivity of this phase is lower by 2 - 4 orders of magnitude as compared to phase-I. NbS$_3$-I exhibits a dielectric behavior with an activation energy of $(3-5) \times 10^3$ K [2,20,21]. Some features of $R(T)$ and the non-linear conductivity [21,22] are not understood yet. In contrast to the transport properties, the crystalline structures of NbS$_3$-I and NbS$_3$-II are similar and closely related. The NbS$_3$-II phase, whose elementary cell contains 8 Nb chains, can be considered as composed of 4 elementary cells of NbS$_3$-I [23] but without dimerization. Only the NbS$_3$-I phase has been previously studied under applied pressure. An insulator-metal transition with an increase of conductivity by 5 orders of magnitude has been reported [24,25] under pressures at about 5 GPa. It is of interest to compare the pressure-induced effects of NbS$_3$-I and NbS$_3$-II phases, due to their contrasting electronic and geometry structures.

In this paper, we report the evolution of Raman spectra under pressure for NbS$_3$-I and NbS$_3$-II. We found that both phases clearly reveal a transition into metallic states at pressures 6 – 7.7 GPa. The dielectric states of both phases re-enter at higher pressures. The qualitatively new high-pressure Raman spectra suggest that both NbS$_3$ phases show a CDW-metal-CDW transition, where the high-pressure re-entrant CDWs show orderings quite different from those of the low-pressure CDWs, probably, directed perpendicular to the chains direction. The isostructural phase transitions in this pressure range are confirmed by the x-ray studies of phase II under pressure.

## II. DETAILS OF EXPERIMENT AND STRUCTURE IDENTIFICATION.

The high-quality whiskers of both phases were synthesized in the V.A. Kotel'nikov Institute of Radio engineering and Electronics of Russian Academy of Sciences [12,26]. High pressure synchrotron XRD experiments were carried out at the 16-BMD station of High-Pressure Collaborative Access Team (HPCAT), Advanced Photon Source, Argonne National Laboratory, with a wavelength of 0.31 Å. High-pressure Raman measurements were done at HPSTAR, China, using the Renishaw Raman spectrometer with a 633 nm excitation source. The sample was packed into a 150 micron diameter hole of a stainless steel-T301 gasket in a diamond anvil cell with 300 micron culets. Silicone oil was used as the pressure-transmitting medium for the Raman scattering experiments, whereas Neon gas was used as the pressure-transmitting media to maintain the hydrostaticity during the high pressure synchrotron powder XRD measurements. A ruby ball was placed in the sample chamber to serve as a pressure caliber using the ruby fluorescence technique [27]. The XRD patterns at various pressures were recorded with a Mar345 imaging plate detector. General Structure Analysis System (GSAS) was employed to resolve the structure and retrieve the lattice parameters. The electronic band structures of NbS$_3$-I under ambient and high pressure were also calculated using the density functional approach (DFT) [28]. DFT calculations were carried out using the Vienna ab-initio Simulation Package (VASP) [29]. A 12 × 9 × 7 Monkhorst-Pack [30] *k*-point mesh was employed for integration of the Brillouin Zone. For the pseudopotentials, the projector-augmented-plane wave (PAW) [31] potentials were used. The Nb and S potentials were considered utilizing the Perdew-Burke-Ernzerhof [32] exchange correlation functional with an energy cutoff of 258 eV.

## III. THE RESULTS

The x-ray powder diffraction pattern for the NbS$_3$-II sample under ambient pressure at room temperature is shown in Fig. 1a, from which the good crystalline data can be fitted to a monoclinic structure with lattice parameters of *a* = 9.84 Å, *b* = 3.39 Å, and *c* = 19.01 Å, with β ≈ 97.4°, close to the reported values of *a* = 9.65 Å, *b* = 3.345 Å, and *c* = 18.75 Å, with β ≈ 98.7° [2,17,33]. The Raman spectrum of NbS$_3$-II at room temperature is shown in Fig. 1b. The incident and scattered light polarizations were parallel to the *b*-axis [34]. The peak positions were obtained using Gaussian fitting. We observed seven dominant peaks at 106, 151, 195, 255, 299, 391 and 571 cm$^{-1}$. NbS$_3$-II can be compared with NbS$_3$-I, whose Raman spectra (Fig. 1c) show a very similar peak structure. Fig. 1c also agrees with the previous works [35]. The observed Raman peaks have been unambiguously assigned [35]: two low-frequency peaks at 106 and 151 cm$^{-1}$ originate from deformation and lattice compression, respectively, along the columns of the Nb-Nb stretching mode; the major peak at 195 cm$^{-1}$ was assigned to (Nb-S$_2^{2-}$); a peak at 255 cm$^{-1}$ originates from the symmetrical valence of the Nb-S vibrations; the peaks at 299 and 391 cm$^{-1}$ appear from the (Nb-S$^{2-}$) mode of intra-chain and inter-chain vibrations, respectively; and the high-wave-number band at 571 cm$^{-1}$ originates from S-S bond stretching vibrations in the (S-S) interaction. As we noted above, the similarity of the ambient-pressure Raman spectra of phases I and II (Fig. 1b,c) is not unexpected due to their structural similarity. However, the majority of the peaks in phase II are coupled with other weaker peaks to form doublets: 106-131, 151-159, 195-201, 255-262, 299-302, 339-349, and 391-405 cm$^{-1}$. This effect can

be due to the positioning of Nb atoms slightly away from the mirror planes. In phase II doubling of the unit cell along the *c*-axis due to CDW modulation has been revealed in STM studies [36].

The in-situ synchrotron XRD data of NbS$_3$-II sample taken under pressures up to 20 GPa are shown in Fig. 2a. Under pressure, the diffraction peaks shift towards higher angle due to the compression of the inter-atomic bonds. It can be also noticed from this figure that the compression rate decreases above 7.66 GPa. GSAS refinement on XRD patterns of NbS$_3$-II under the pressures of 5 and 12 GPa are shown in Fig. 2b(i-ii) respectively. The lattice parameters are retrieved for each pressure, and the pressure dependencies of lattice parameters *a*, *b*, and *c* are shown in Fig. 3a; $\beta$ vs. pressure is shown in the inset to Fig. 3a. The external pressure affects the lattice parameters through a reduction in all axes and an increase in the oblique intersection angle. However, the rate of change in all parameters differs below and above 7.7 GPa. Fig. 3b illustrates the pressure dependence of the normalized lattice parameters based on the Rietveld refinements. The normalized lattice parameters *a*, *b* and *c* contract by 4.9%, 3.3% and 3.2%, exhibiting anisotropic compression. This effect is likely to reveal an electronic phase transition in NbS$_3$-II. The change in the unit cell volume with pressure is shown in Fig. 3c. The third-order Birch-Murnaghan equation of state (EOS) [37] has been fit separately for two regions, below and above 7.7 GPa (Fig. 3c):

$$P = \frac{3B_0}{2}\left[\left(\frac{V_0}{V}\right)^{7/2} - \left(\frac{V_0}{V}\right)^{5/2}\right]\left\{1 + \frac{3}{4}(B\_0' - 4)\left[\left(\frac{V_0}{V}\right)^{2/3} - 1\right]\right\}$$

where $V_0$ is the unit cell volume at ambient pressure, $B_0$ is the bulk modulus at ambient conditions, and $B'_0$ is its pressure derivative. The fittings yield $B_0 = 70$ GPa below 7.7 GPa and $B_0 = 122$ GPa above 7.7 GPa, close to the theoretically reported value of 117 GPa(see supplementary information of [38]). In the pressure-released state (0.02 GPa), all peaks are almost reverted to their ambient XRD pattern; the relaxed mode of atoms shows the elasticity of NbS$_3$-II, although the intensities are not restored.

Fig. 4a shows the Raman spectra of NbS$_3$-II under hydrostatic pressure up to 15 GPa. The diamond background of each data point was subtracted by base line fittings. The peaks' positions vs. pressure are given in Fig. 4b. All the modes shift gradually to higher wave numbers, up to 5.9 GPa, and no other changes can be seen up to 6 GPa. Upon further increase of pressure, all the peaks decrease or almost vanish in the range 6.1 – 7.9 GPa. The Raman peaks can be seen again starting from 8.6 GPa. However, they appear at different wave-number positions. Beginning from 8.6 GPa, the modes from the same elemental sites such as Nb-Nb (151 cm$^{-1}$) and S-S (571 cm$^{-1}$) shift to lower wavenumbers, whereas the vibration modes between the atoms of different elements (between Nb-S) shift to higher wave numbers. This means that Nb-S bonds become more compressed, and the vibration mode frequencies shift upwards under pressure. The wavenumbers of the NbS$_3$-II optical phonons modes observed at two different regions are listed in the Table I.

The disappearance of Raman peaks around 7 GPa, in the gray-color region (Fig. 4b), can be attributed to a transformation of NbS$_3$-II into a metallic state. The direct observation of dielectric-metal transition in phase I under pressure [23], as well as the similarity of the ambient Raman spectra of phases I and II, allows us to suggest that the behavior of the two phases under pressure is similar. This inspired us to study the pressure evolution of the Raman shifts in phase I (Fig. 5a). The result appears very similar with that of phase II. At 6.5 GPa the peaks become nearly invisible, and then, at 7.7 GPa, they appear at new positions. The plots of the peaks' positions *vs.* pressure for phases I and II look very much alike (Figs. 4b and 5b). From this we can conclude that under pressure the properties of phases I and II evolve in similar ways. Both phases, after passing a critical region around 7–8 GPa, show the formation of new vibration peaks distinctly different from the low-pressure ones.

## IV. DISCUSSION

In the metallic state the penetration depth of laser irradiation decreases drastically, giving rise to the extinguishing of the Raman peaks, and the phonon vibration with a metallic background is more damped. In addition, suppression of the CDWs increases the symmetry of the lattice and could deactivate some peaks. For instance, a clear evidence for the tight coupling between the CDW condensate and the vibration modes (at lower frequencies, though) is observed in rare-earth tri-tellurides, revealed by their pressure and chemical pressure dependences [39]. One can see that formation of a CDW could result in an enrichment of the spectra. In ref. [40] the Raman shift and the fast Fourier transform of the transient oscillations, following fast suppression of CDW in $K_{0.3}MoO_3$, were confronted, giving a further example of CDW manifestations in Raman shift (see also [41,42,43,44]). Though we cannot distinguish the contributions of the penetration depth fall and of the vibration modes deactivation, it is clear that in the "gray" pressure areas (Figs. 4b, 5b) the CDWs become suppressed. Then it is reasonable to suppose, that above the critical region the CDWs recover with different orderings, while the lattice in itself does not suffer phase transformations as it can be seen from the x-ray spectra of $NbS_3$-II. The number of Raman peaks is likely to remain the same, though some of them could enter or go out of the detection frequency window.

The initial suppression of the CDWs correlates with the increase of inter-chain hopping, suppression of the one-dimensionality of $NbS_3$ and, thus, the impairing of the Fermi surface nesting conditions. In this sense, the effect of pressure is analogous to that of uniaxial strain, $\varepsilon$. For $NbS_3$-II [12], as well as for the relative compounds orthorhombic $TaS_3$ and $NbSe_3$ [11,45,46], elongation of the whiskers suppresses the CDW transitions reducing $T_P$. One can assume that the behavior of orthorhombic $TaS_3$ and $NbS_3$-II under mechanical impacts is roughly similar. In fact, for $NbS_3$-II $dT_{P1}/d\varepsilon \approx -5500$ K [12], which is close to the values of $dT_P/d\varepsilon$ in orthorhombic $TaS_3$ [11,45]. In $TaS_3$ the CDW transition is completely suppressed at P~10 GPa. Thus, it is reasonable to suppose, that the pressure of the same order will suppress CDW-1 (or/and CDW-0) in $NbS_3$-II, or, at least, reduce $T_{P1}$ (or/and $T_{P0}$) below room temperature.

Pressure is a very clean tool[74-53], one can assume that above the critical region the pressure can provoke formation of a new CDW directed normally to the conducting chains. This idea is supported by the features of a related compound $ZrTe_3$, where the CDW is directed perpendicular to the *b*-axis and is stimulated by pressures up to 2 GPa [5]. Some features of the high-*P* Raman spectra could be attributed to activation of vibration modes perpendicular to the *b*-axis, also in favor of the analogy with $ZrTe_3$.

The transition of the metallicity can also be examined by the Fermi surface evolution under pressure. The calculated electronic band structures for $NbS_3$-I at 0 and 6 GPa are shown in Figs. 6a and 6b, respectively. The pictorial representation of Fermi surface at 5 and 6 GPa are given Figs.7a, 7b respectively. At 6 GPa only one band crosses the Fermi level; however a slight further compression drastically transforms the topology of the Fermi surface. More bands are crossing the Fermi level, thereby more Fermi surfaces evolve, thus taking the system through considerable Lifshitz transitions. As a result, the electronic density of states becomes quite high at the Fermi level, making the system a good metal. The metallization at approximately 6-7 GPa is manifested in the Raman spectrum, where the peaks vanish. Since, the Raman spectra of both phases show similar evolutions, similar Lifshitz transitions can be assumed for the $NbS_3$-II as well.

## V. CONCLUSION

In summary, the pressure effect on the monoclinic $NbS_3$ has been studied with XRD and Raman techniques. The x-ray studies show no structural phase change up to 20 GPa, but the rate of lattice compression jumps down at 7.5 – 8 GPa indicating a possible isostructural electronic phase transition. The transitions (or a two-stage transition) are clearly revealed by the Raman-shift spectra. Namely, they demonstrate: (i) disappearance of the Raman peaks at about 7 GPa suggesting a suppression of the CDWs and formation of a metallic state in this pressure range, and (ii) recovery of Raman peaks at about 8.6 GPa, but at quite different frequencies. This result reveals a change of the electronic environment of the Nb and S atoms and suggests re-entrance of the CDW above 8 GPa. Very similar Raman spectra (see also [34,35]) and their pressure evolution are found for the triclinic $NbS_3$. The observation of the insulator-metal transition under pressure for this phase [14, 15] justifies the above interpretation and suggests similarity of the behaviors of phases I and II $NbS_3$ under pressure. The results can be attributed to the increase of interchain coupling under pressure, which impairs the Fermi surface nesting and leads to the destruction of the CDW phase at pressures above 7.66 GPa. At higher pressures a new ordering appears, probably, directed perpendicular to the *b*-axis, similar to the case of $ZrTe_3$.

**ACKNOWLEDGEMENT**

We are grateful to V.F. Nasretdinova for the help in sample preparation and useful discussions. This work was financially supported by the National Nature Science Foundation of China (Grant No. 51527801, U1530402). MAH and ANV have been supported by the Ministry of Education and Science of the Russian Federation in the framework of Increase Competitiveness Program of NUST "MISiS" grant K2-2017-084, by acts 211 of the Government of Russian Federation, Contracts No. 02.A03.21.0004 and 02.A03.21.0011. The support of RFBR (Grant 17-02-01343) is acknowledged. The synthesis of the samples and their preparation for Raman studies was supported by the Russian Scientific Foundation (grant 17-12-01519), V.V. Pavlovskiy contributed to the analysis of ab initio simulations within the framework of the State task. W. W. Pai is supported by MOST, Taiwan (107-2112-M-002 -017 -MY3), and by the AI-MAT center of National Taiwan University (NTU-107L900802).

————

**Figure**

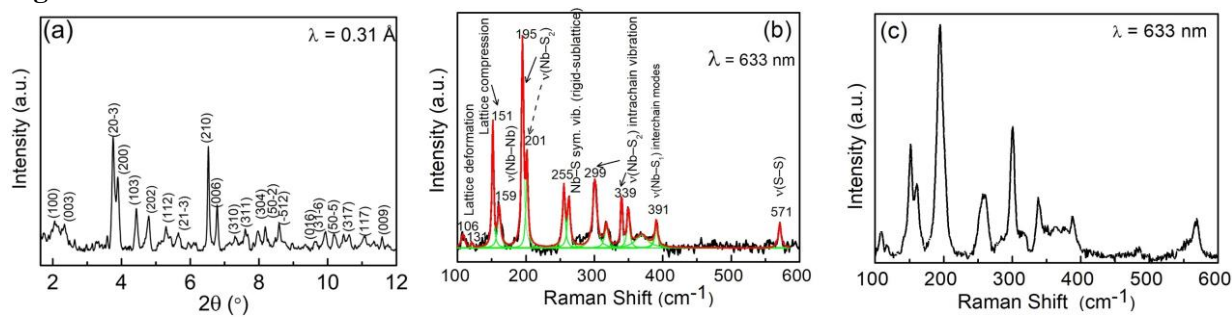

FIG. 1: (a) XRD pattern of NbS$_3$-II; (b) Raman scattering spectrum for NbS$_3$-II with peak and respective vibrational modes [Black: Observed data; Red: Peak fitting; Green: Individual fitted peaks]. (c) Raman scattering spectrum for NbS$_3$-I sample.

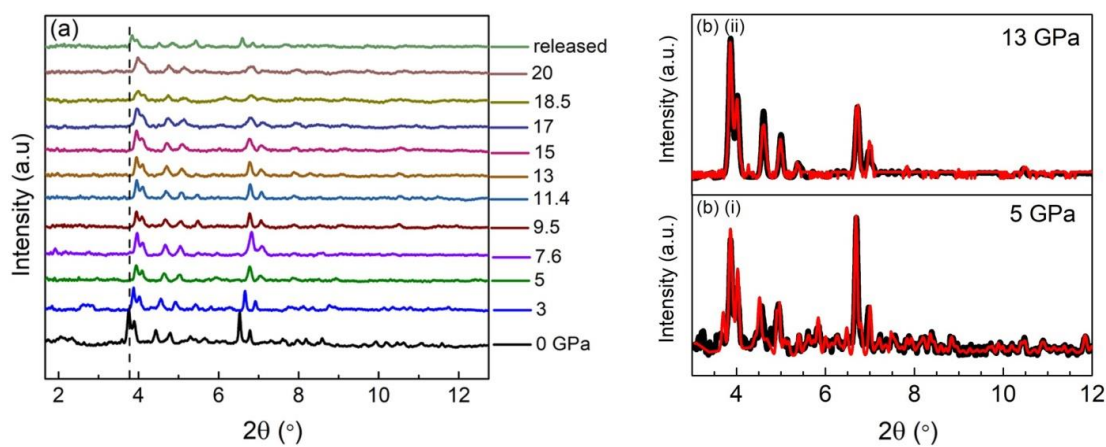

FIG. 2: (a) Synchrotron XRD patterns of NbS3-II under various high pressures (The dashed line is the guide for an eye.); (b) GSAS refinement on XRD patterns of NbS3-II under the pressures of 5 and 13 GPa

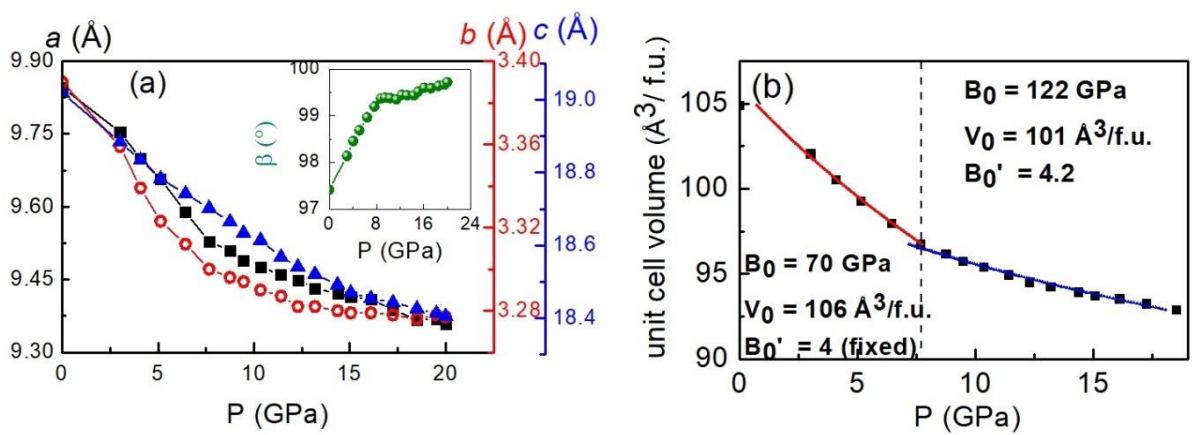

FIG. 3: (a) Pressure dependence of lattice parameters (Left axis: *a*), (first Right axis: *b*) and(second Right axis: *c*) [Inset: Pressure dependence of *β*]; (b) Experimental compression curve of NbS$_3$-II and EoS fitting in separate regions below and above 7.66 GPa, the estimated bulk modulus.

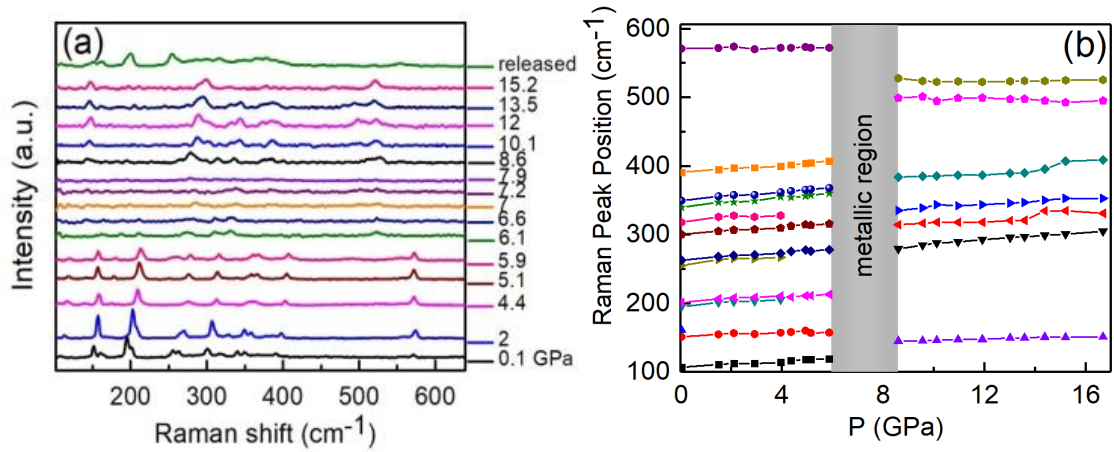

FIG. 4: (a) Raman scattering spectra of NbS$_3$-II under selected pressures up to 15 GPa; (b) Pressure evolution of Raman shifts [Gray color region: The metallic region where is no observance of Raman peaks].

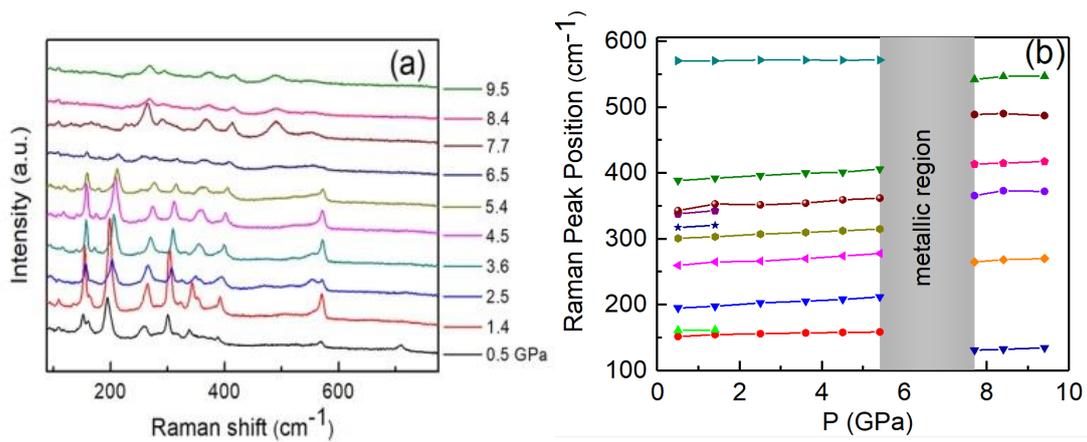

FIG. 5: (a) Raman scattering spectra of NbS$_3$-I under selected pressures up to 9.5 GPa; (b) Pressure evolution of Raman shifts [Gray color region: The metallic region where is no observance of Raman peaks].

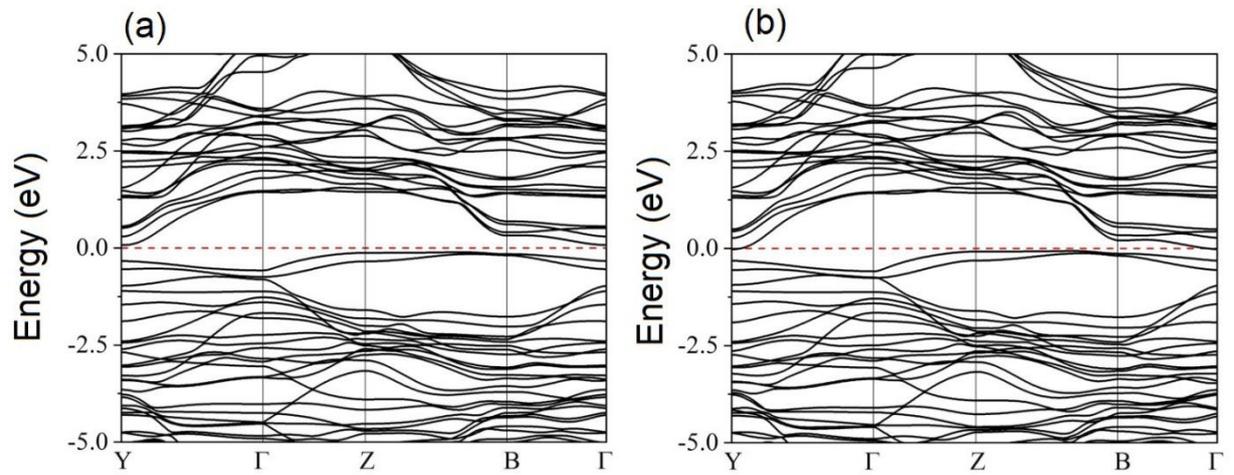

FIG. 6: Electronic band structure of NbS$_3$-I at (a) 0 GPa and (b) 6 GPa.

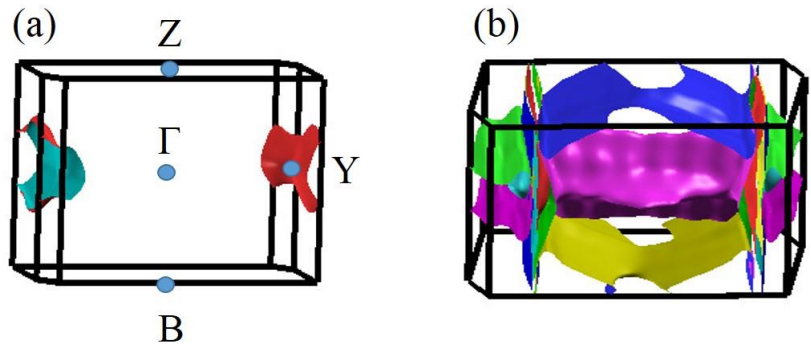

FIG. 7: Fermi surface of NbS$_3$-I at (a) 5 GPa and (b) 6 GPa.

Table I: Assignment of Raman vibrational modes under two different pressure ranges of CDW state such as 0-5.9 GPa and 8.6-15.2 GPa, for the NbS$_3$-II.

| vibrational modes | Deformation | Lattice compression and ν(Nb–Nb) | ν (Nb–S$_2$) | Nb–S sym. vib. (rigid-sublattice) | ν (Nb–S$_2$) intra-chain vibration | ν (Nb–S$_2$) intra-chain vibration | ν (Nb–S$_1$) inter-chain modes | ν (S–S) | Ref. |
|---|---|---|---|---|---|---|---|---|---|
| 0 - 5.9 GPa | 106-131 | 151-159 | 195-201 | 255-262 | 299-302 | 339-349 | 391-405 | 571 | [28, 29] |
| 8.60 - 15.2 GPa | | 145.1 | 279.5 | 314.7 | 335 | 384 | 499 | 528 | |